\documentclass[iop]{emulateapj}

\usepackage{graphicx}

\newcommand{\sj}{XTE J1701--462}
\newcommand{\ledd}{$L_{{\rm Edd}}$}
\newcommand{\mdot}{$\dot{M}$}
\newcommand{\mdisk}{$\dot{M}_{d}$}
\newcommand{\xte}{{\it RXTE}}

\shorttitle{\sj}
\shortauthors{Jeroen Homan et al.}

\begin{document}

\title{XTE J1701--462 and its Implications for the Nature of Subclasses in Low-Magnetic-Field Neutron Star Low-Mass X-Ray Binaries}

\author{
Jeroen Homan\altaffilmark{1},
Michiel van der Klis\altaffilmark{2},
Joel K.\ Fridriksson\altaffilmark{1},
Ronald A. Remillard\altaffilmark{1},
Rudy Wijnands\altaffilmark{2},
Mariano M\'endez\altaffilmark{3},
Dacheng Lin\altaffilmark{4},
Diego Altamirano\altaffilmark{2},
Piergiorgio Casella\altaffilmark{5},
Tomaso Belloni\altaffilmark{6},
and Walter H.\ G.\ Lewin\altaffilmark{1}
}

\altaffiltext{1}{MIT Kavli Institute for Astrophysics and Space
Research, 70 Vassar Street, Cambridge, MA 02139;
jeroen@space.mit.edu}

\altaffiltext{2}{Astronomical Institute `Anton Pannekoek', University of Amsterdam, Science Park 904, 1098 XH, Amsterdam, The Netherlands}

\altaffiltext{3}{Kapteyn Astronomical Institute, Groningen University, 9700 
AV, Groningen, The Netherlands}

\altaffiltext{4}{Centre d'Etude Spatiale des Rayonnements, UMR 5187, 9 av.\ du Colonel Roche, BP 44346,
31028 Toulouse Cedex 4, France}

\altaffiltext{5}{School of Physics and Astronomy, University of Southampton, Southampton, Hampshire, SO17 1BJ, United Kingdom}

\altaffiltext{6}{INAF-Osservatorio Astronomico di Brera, Via E.
Bianchi 46, I-23807 Merate (LC), Italy}

\begin{abstract}

We report on an analysis of RXTE data of the transient neutron star
low-mass X-ray binary (NS-LMXB) XTE J1701-462, obtained during its
2006-2007 outburst. The X-ray properties of the source changed between those of various types of NS-LMXB subclasses. At high
luminosities the source switched between two types of Z source behavior
and at low luminosities we observed a transition from Z source to atoll
source behavior. These transitions between subclasses primarily manifest
themselves as changes in the shapes of the tracks in X-ray color-color
and hardness-intensity diagrams, but they are accompanied by changes in
the kHz quasi-periodic oscillations, broad-band variability, burst
behavior, and/or X-ray spectra. We find that the low-energy X-ray flux
is a good parameter to track the gradual evolution of the tracks in color-color and hardness-intensity diagrams, allowing us to resolve the
evolution of the source in greater detail than before and relate the
observed properties to other NS-LMXBs. We further find that during the
transition from Z to atoll, characteristic behavior known as the atoll
upper banana can equivalently be described as the final stage of a
weakening Z source flaring branch, thereby blurring the line between the
two subclasses. Our findings strongly suggest that the wide variety in
behavior observed in NS-LXMBs with different luminosities can be linked
through changes in a single variable parameter, namely the mass
accretion rate, without the need for additional differences in the
neutron star parameters or viewing angle. We briefly discuss the
implications of our findings for the spectral changes observed in NS
LMXBs and suggest that, contrary to what is often assumed, the position
along the color-color tracks of Z sources is not determined by the
instantaneous mass accretion rate.

\end{abstract}


\section{Introduction}\label{sec:intro}

Neutron star low-mass X-ray binaries (NS-LMXBs) are systems in which a low-magnetic-field neutron star
accretes matter from a low-mass companion star through Roche-lobe overflow. They display a wide range of X-ray spectral and variability properties, both as a class and as individual sources \citep[see][for a recent review]{va2006}. Based on their correlated X-ray spectral and rapid variability behavior, NS-LMXBs are commonly divided into two subclasses, the so-called  Z sources and the atoll sources \citep{hava1989}, named after the shapes they trace out in X-ray color-color (CD) and hardness-intensity diagrams (HID).  The Z sources have luminosities close to the Eddington luminosity ($\sim$0.5--1\,\ledd\ or more), whereas atoll sources have luminosities in the range $\sim$0.01--0.5\,\ledd. 

\begin{figure}[h]
\vspace{0.5cm}
\epsscale{0.75}
\plotone{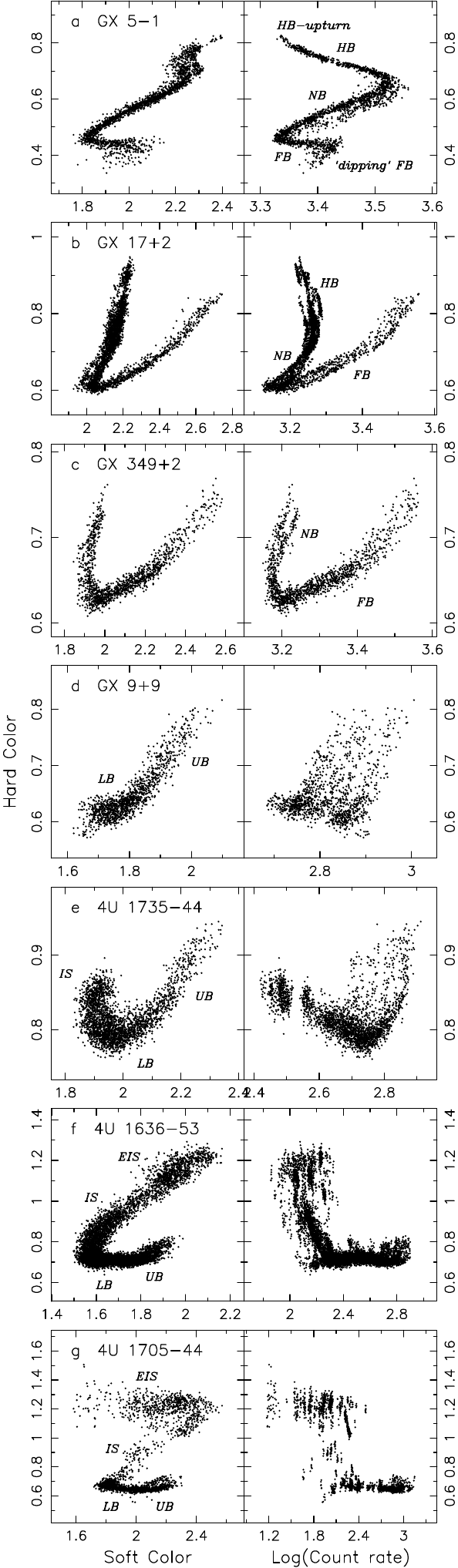}
\caption{Example CDs (left) and HIDs (right) of the Z source NS-LMXBs GX 5--1, GX 17+2, and GX 349+2, and the atoll source NS-LMXBs GX 9+9, 4U 1735--44, 4U 1636--53, and 4U 1705--44. Each data point represents a 256s average. The diagrams are taken from a compilation by Fridriksson et al.\ (2010, in prep.) and were created following the steps described and referred to in \S\ref{sec:data} and \S\ref{sec:results}. Z and atoll branches/states are indicated in the CD or HID of each source. They are: horizontal branch (HB), normal branch (NB), flaring branch (FB), upper-banana branch (UB), lower-banana branch (LB), island state (IS), and extreme island state (EIS). For the Z source GX 5--1 we have also labelled the upturn from the horizontal branch (HB-upturn), and the `dipping' flaring branch (dipping FB).}
\label{fig:overview}
\end{figure}

An example of the variety in shapes that is observed in the CD/HID tracks of the Z and atoll subclasses is shown in Figure \ref{fig:overview}. The Z source tracks typically show three branches, which from top to bottom are called the horizontal branch, the normal branch, and the flaring branch. Based on the presence and orientation of these branches, the Z sources can be further divided into the `Cyg-like' Z sources (Cyg X-2, GX 5--1, and GX 340+0) and the `Sco-like' Z sources (Sco X-1, GX 17+2, and GX 349+2), with the former showing `Z'-shaped tracks in the HID (Fig.\ \ref{fig:overview}a) and the latter more `$\nu$'-shaped tracks (Fig.\ \ref{fig:overview}b,c; see also \citealt{kuvaoo1994,hovawi2007} (hereafter H07)). The atoll sources can show `Z'-shaped tracks as well (Fig.\,\ref{fig:overview}g), although the individual branches are thought to have a different physical nature than the Z source branches \citep{baol2002,vavame2003,revava2004,va2006}. The branches of the Z-shaped atoll tracks are called, from top to bottom (Fig.\ \ref{fig:overview}g), the extreme island state, the island state, and the lower and upper banana branches; these branches are also frequently referred to as the hard, transitional/intermediate, and soft states, respectively. The low-luminosity atolls are typically found in the (extreme) island state, whereas the high-luminosity ones tend to be found mainly on the (upper) banana branch (Fig.\ \ref{fig:overview}d).

 In addition to motion along atoll or Z branches, some sources also show changes in the shape and location of their tracks in the CD and HID, which is often referred to as secular motion. Examples of this can be seen in Figure \ref{fig:overview}, with the horizontal branch of GX 17+2 and the upper-banana branches of GX 9+9 and 4U 1735--44 being traced out at varying intensity levels. Much stronger secular motion can be found in the Z source Cyg X-2 \citep{kuvava1996,wivaku1997}, where it occasionally results in switches between Cyg-like and Sco-like Z source behavior \citep[see Figure 1 in][]{murech2002}. Finally, we note that transitions between the atoll (extreme) island state and the banana branches, and vice versa, often show strong hysteresis in HIDs, resulting in jumps between those branches  \citep[e.g.,][see also Fig.\ \ref{fig:overview}f,g]{maba2004,gldogi2007,behomo2007}. It is not clear if and how this behavior is related to the secular motion.

In addition to differences in X-ray luminositiy and CD/HID tracks, the NS-LMXBs subclasses also show differences in X-ray spectral and variability properties \citep{va2006}, radio behavior \citep{mife2006}, as well as in the rate at which they show type I (thermonuclear) X-ray bursts \citep{gamuha2008}. This disparity in the properties between the subclasses was a long-standing problem in the understanding of the accretion processes in NS-LMXBs. While differences in the mass 
accretion rate (\mdot) could explain some of the variety in observed properties (e.g., in 
X-ray luminosity and bursting behavior),  it was unclear if one could link the wide variety in CD/HID tracks, rapid variability, and broadband spectral properties through changes in \mdot\ alone.  
Additional parameters have been proposed to explain some of the 
observed differences, not only between Z and atoll subclasses, but also between the Cyg-like and Sco-like Z sources: magnetic field strength and 
viewing angle \citep[e.g.,][]{pslami1995,hava1989,kuva1995}.

Testing for the necessity of such additional parameters is complicated by the fact that none of the classical Z sources has been observed to venture into the atoll luminosity range (and vice versa). Until recently only two sources were suspected to have made a transition between Z-like and atoll-like behavior: Cir X-1 and XTE J1806--246 (= 2S 1803--245). Cir X-1 was found to show Z-like properties at high luminosities \citep{shbrle1999}  and atoll-like properties at low luminosities \citep{oovaku1995}. XTE J1806--246 showed some  characteristics of Z sources near the peak of its outburst, which reached $\sim$\ledd\  (other bright NS-LMXB transients typically only reach $\sim$0.5\,\ledd), and atoll properties at lower luminosities \citep{wiva1999b}.  While these two examples suggested that a single source can show atoll/Z source characteristics depending on luminosity, and lent some credence to the belief that the NS-LMXB subclasses are entirely the result of differences in \mdot, the actual evolution from Z tracks to atoll tracks had never been observed. 

Significant progress on the issue of NS-LMXB subclasses was gained following the outburst of \sj, a transient NS-LMXB that was discovered in January 2006 \citep{reli2006} and remained in outburst for $\sim$19 months. Initial observations with the {\it Rossi X-ray Timing 
Explorer (RXTE)} revealed that \sj\ was the first transient NS-LMXB to show 
all of the Z source characteristics in X-rays (H07). The radio properties of the source were consistent with those of the Z sources as well \citep{fedaho2007}. Observations of type I X-ray bursts with radius expansion, near the end of the outburst, put the source at a distance of 8.8$\pm$1.3 kpc and indicate that during the early phase of the outburst the source had super-Eddington luminosities \citep{lialho2009}. After the early phase of the outburst the source made a transition from  Cyg-like to Sco-like Z source behavior as the source intensity dropped by a factor of 1.5 (H07), and observations near the end of the outburst, in July 2007, showed the source evolving from Sco-like Z source behavior to atoll source behavior \citep{hobewi2007,howial2007}.  A full spectral analysis of the entire outburst  was performed by \citet[hereafter LRH09]{lireho2009}, who confirm that at the lowest luminosities the spectral behavior of \sj\ is consistent with that observed for the two atoll source transients Aql X-1 and 4U 1608--52 \citep{lireho2007}. \sj\  is the first source in which the evolution from Z to atoll source has been observed in detail, and its behavior strongly suggests that different levels of \mdot\ are indeed sufficient to explain the different NS-LMXB subclasses.

The outburst of \sj\ also allows for a study of the evolution of the X-ray properties along CD/HID tracks within the context of the different NS-LMXB subclasses. This may shed new light on the nature of the Z and atoll branches, whether some or all are unique to each class, and, most importantly, what drives the motion along them. As far as atoll sources are concerned, past observations of transient NS-LMXBs have suggested that the evolution along atoll source CD/HID tracks is mainly driven by variations in \mdot\ \citep[see, e.g.,][]{va1994b,lireho2007}, increasing from the extreme island state to the upper banana branch. However, the aforementioned hysteresis in spectral state transitions \citep{maco2003} and  parallel tracks observed in HIDs \citep[see, e.g.,][]{mevafo2001,dimeva2003} hint at a more complex relation between \mdot\ and spectral state in atoll sources \citep{li2009}, not unlike what is suggested by the spectral evolution seen in transient black-hole X-ray binaries \citep{hobe2005}.

It should be noted that there is 
ambiguity in some of these discussions about the definition of \mdot.  Some
authors implicity define \mdot\ as the mass transfer rate from the companion to the 
compact object Roche lobe, while others take the mass flow rate through the 
inner disk or the accretion rate onto the neutron star surface as the more relevant
definition. Others explicitly distinguish various \mdot\ components.  In the presence 
of the various proposed flows in the accretion/ejection process, such as disk and 
equatorial boundary layer flows, spherical inflows, magnetically dominated polar flows, 
possible disk winds and jets, clearly this is a complex issue to which we shall 
return in \S\ref{sec:discussion}.

Interpreting the evolution along the Z source tracks has been more difficult, owing to the fact that motion along the Z source tracks corresponds to relatively small (less than a factor of $\sim$2) and non-monotonic intensity variations. For Z sources it was suggested that \mdot\ increases from the horizontal branch to the flaring branch \citep{havaeb1990}, although other scenarios have been put forward as well: e.g., \mdot\ changing in the opposite direction \citep{chhaba2006}, or \mdot\ not changing at all \citep{hovajo2002}. Based on their spectral analysis of the Z tracks, LRH09 suggest that motion along the Z tracks may occur at a nearly constant \mdot, with the branches being the result of different instabilities in the near-\ledd\ accretion flow.

More detailed analyses of the data presented in this work, dealing with
different aspects of \sj, can be found in a number of other 
papers: spectral analysis (LRH09), type-I X-ray 
bursts \citep{lialho2009}, kHz QPOs \citep{sameal2010}, broad-band 
variability (Aresu et al.\ 2010, in prep.), and the rapid decay into quiescence \citep{frhowi2010}. 
Here, we present an overview of the outburst of \sj\ that focuses on aspects of the evolution of \sj\ in which the change from Z source to atoll source is most clearly seen: the CD/HID tracks, high-frequency QPOs, and broad-band variability.  In particular, we find that selections of data groups based on low-energy count rate allow for a more detailed study of the evolution of the CD/HID tracks than in H07 and LRH09. This also permits a more precise comparison of the observed tracks for \sj\ with those found in the various NS-LMXB sub-classes. In \S\ref{sec:data} we summarize our data set and analysis techniques. Our results are presented in \S\ref{sec:results} and in \S\ref{sec:discussion} these results are interpreted and discussed within the framework of various scenarios for the role of \mdot\ in driving the evolution between NS-LMXB subclasses and along the CD/HID tracks of NS-LMXBs.

\section{Observations and Data Analysis}\label{sec:data}

We analyzed all 865 pointed \xte\ \citep{brrosw1993} observations of \sj\ made between 2006 January 19 (MJD 53754) and 2007 Aug 29 (MJD 54341). Five of them were discarded from further analysis because proportional counter unit 2 (PCU2) was not working or because the length of usable data intervals was too short ($<$256s) for our variability analysis. The remaining 860 observations had a total combined exposure time of $\sim$2.73 Ms. For our analysis we only made use of data from the Proportional Counter Array \citep[PCA;][]{jamara2006}. We followed the same analysis steps as described in H07, the only difference being that in addition to background corrections, the {\tt standard-2} data used for the light curves and CDs/HIDs in \S\ref{sec:lc-cd} were also corrected for dead time. All count rates and colors given in the text and used in the figures are for PCU2 only; for our power spectral analysis we made use of all active PCUs. Dates will be referred to as days since 2006 January 19 (MJD 53754).

\section{Results}\label{sec:results}

\subsection{Light curves and color-color diagrams}\label{sec:lc-cd}

\begin{figure*}[tbp]
\epsscale{0.8}
\plotone{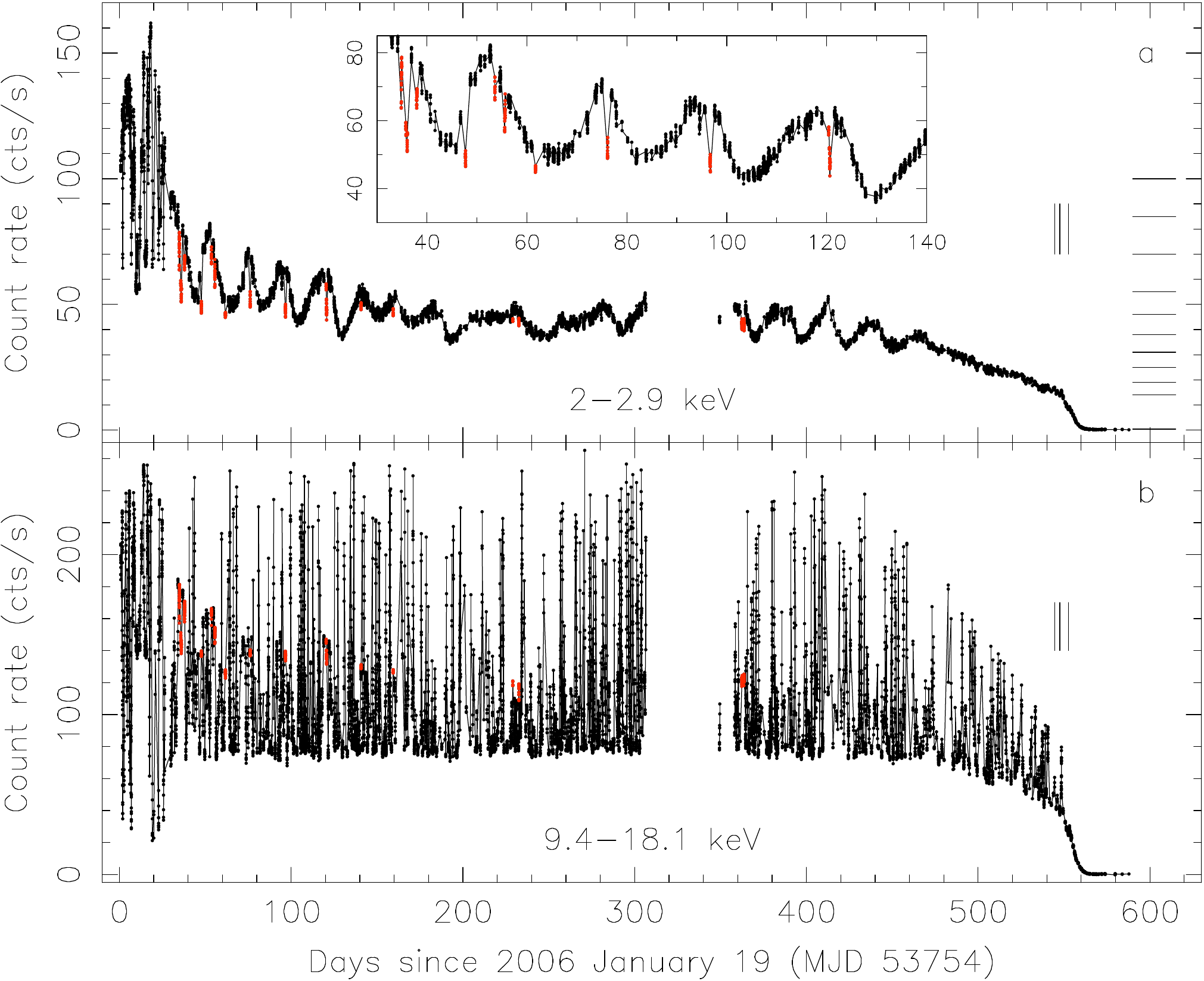}
\caption{{\it RXTE}/PCA light curves of \sj\ in two energy bands; $\sim$2--2.9 keV (top) and $\sim$9.4--18.1 keV (bottom). The count rates are from the PCU2 detector and the time resolution of the light curves is 256 s. Error bars are smaller than the plot symbols. The inset in the top panel zooms in on the first few long-term cycles and the vertical lines near day 550 represent the start times of the three type I X-ray bursts that were detected.  The count rate levels used for our CD/HID selection process, which is described in \S \ref{sec:lc-cd}, are shown on the right-hand side of the top panel. Red data points are from observations (on the Z source horizontal branch) that were moved to a different CD/HID in this selection process.}
\label{fig:lc}
\end{figure*}

In Figure \ref{fig:lc} we plot {\it RXTE}/PCA light curves of \sj\ in two energy bands: $\sim$2--2.9 keV and  $\sim$9.4--18.1 keV, corresponding to {\tt standard-2} channels 1--3 and 20--40, respectively. The two light curves have a strikingly different appearance. Except for the first 30 days of the outburst and occasional short drops in intensity between days 30 and 130 (see inset in Fig.\ \ref{fig:lc}a), the low-energy light curve shows little short-term variability. Smooth long-term modulations with periods of $\sim$20--50 days are present, until the source starts its descent into quiescence. The high-energy light curve, on the other hand, shows strong short-term flaring, which, as we discuss later, corresponds mostly to motion along the Z source flaring branch. 
This flaring starts to weaken around the time at which the long-term low-energy modulations end, and it abruptly subsides when the decay at low energies accelerates (day $\sim$550). After day $\sim$560, the count rates reach a very low, but 
non-zero, level of $\sim$2 cts\,s$^{-1}$ per PCU (full energy band). This residual emission can be attributed to diffuse Galactic emission \citep{frhowi2010}, but has not been subtracted from our data. The times of the three type I X-ray bursts that were detected near the end of the outburst \citep{lialho2009} are indicated by the short vertical lines in Figure \ref{fig:lc}.

\begin{figure*}[t]
\epsscale{1.0}
\plotone{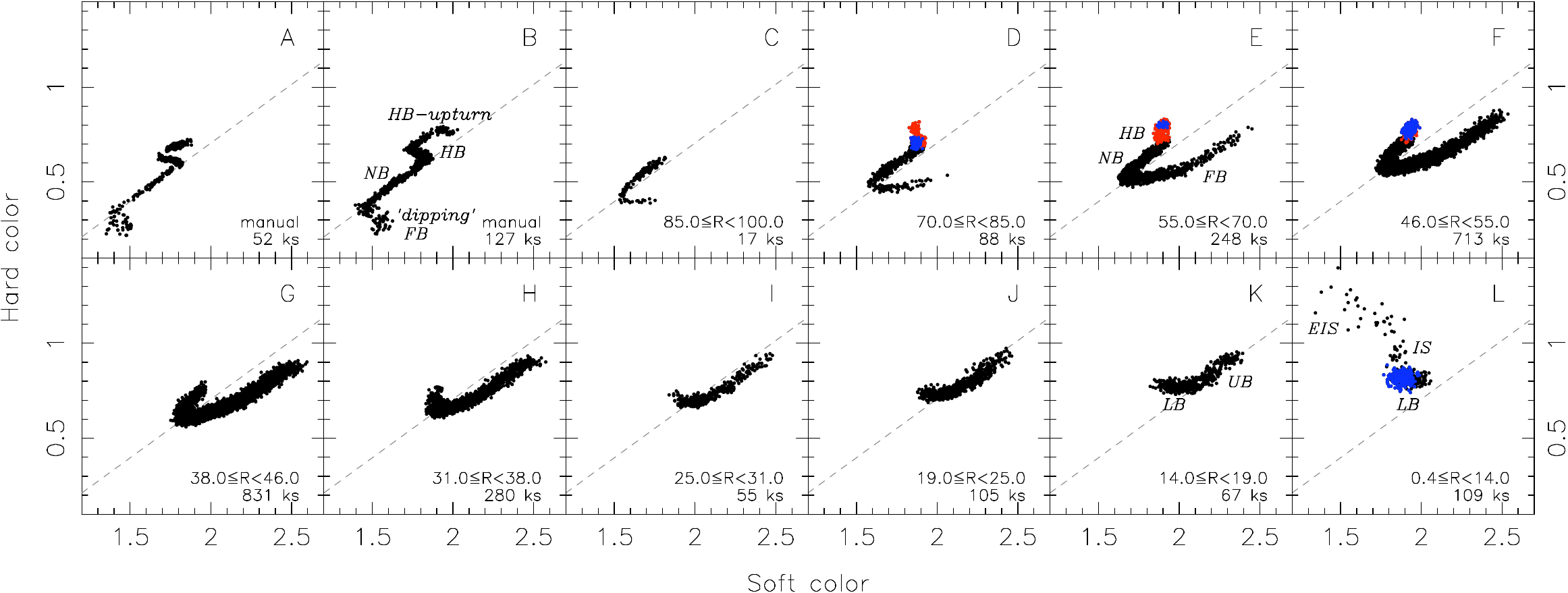}
\caption{The evolution of the CD tracks of \sj\ as a function of low-energy count rate. The selection criteria for the tracks are given at the bottom of each panel, together with the total amount of exposure time. Diagrams C-L are based on selections of the low-energy ($\sim$2.0--2.9 keV) count rate $R$; diagrams A and B are based on manual selections (before day 29).  Soft color and hard color are defined as the ratio of count rates (extracted from {\tt standard-2} data) in  the
$\sim$4.0--7.3 keV (channels 7--14) and $\sim$2.4--4.0  keV  (ch.\ 3--6) bands, and the  $\sim$9.8--18.1 keV (ch.\ 21--40) and $\sim$7.3--9.8 keV (ch.\ 15--20) bands, respectively. Data points represent 64 s averages for high count rates, increasing to 1024 s averages for the lowest count rates. The blue points indicate observations in which kHz QPOs were detected; the red points represent data from observations on the Z source horizontal branch that were moved to the selections corresponding to neighboring non-horizontal branch observations (see \S\ref{sec:lc-cd}). The dashed gray line indicates the approximate path followed by the normal/flaring branch vertex during the transition from Cyg-like Z to Sco-like Z and, finally, atoll source behavior.}
\label{fig:cd}
\end{figure*}

As shown by H07 and LRH09, \sj\ displays strong secular changes in its CD and HID. Creating a single CD/HID for the entire outburst results in a complicated set of overlapping tracks with varying shapes (see, e.g., Figure 6 in LRH09). Time-based selections, such as the ones made by H07 and LRH09, also encounter strong secular motion associated with the 20--50 day modulations, which causes the shape and position of the tracks to change on a time scale of a few days. Fortunately, we have found that the low-energy count rate (2--2.9 keV; see Fig.\ \ref{fig:lc}a) appears to be a fairly good tracer of the secular changes. For most of the outburst, the count rate in this band changes very little as the source traces out a particular Z track (although it does change along the atoll track we observed at the end of the outburst). Furthermore, the low-energy count rate can be used to make separate CDs and HIDs based on count rate intervals selected over most of the outburst. 

Eleven low-energy count rate intervals were defined to group the CD/HID data points; the maximum values of these intervals are shown as horizontal lines on the right-hand side of Figure \ref{fig:lc}a. Although the goal of these selections is to follow the secular evolution of the CD/HID tracks, there are two cases where changes in the low-energy count rate were mainly the result of motion along the CD/HID track. The first case concerns the period before day 29, during the Cyg-like Z source stage of the outburst, when the source shows large intensity swings in the low energy band, associated with motion along the Z. For this period we manually selected data to create two sets of CDs/HIDs (bringing the number of CD/HID selections to a total of thirteen); the first period consists of data from days 14.0--23.0, corresponding to the highest observed count rates, and the second combines data from days 0--14.0 and 23.0--29.0. The second case where we experienced difficulties with the low-energy countrate selections consists of several occasions when the source enters the horizontal branch {\it after} day 29. These instances are marked in red in Figure \ref{fig:lc} and many are clearly visible as drops in count rate near the maxima of the long-term modulations, as the result of which they end up in a lower count rate selection. To correct for this, these data points were grouped with their neighbouring non-horizontal branch observations (typically one count rate level higher).

\begin{figure*}[t]
\epsscale{1.0}
\plotone{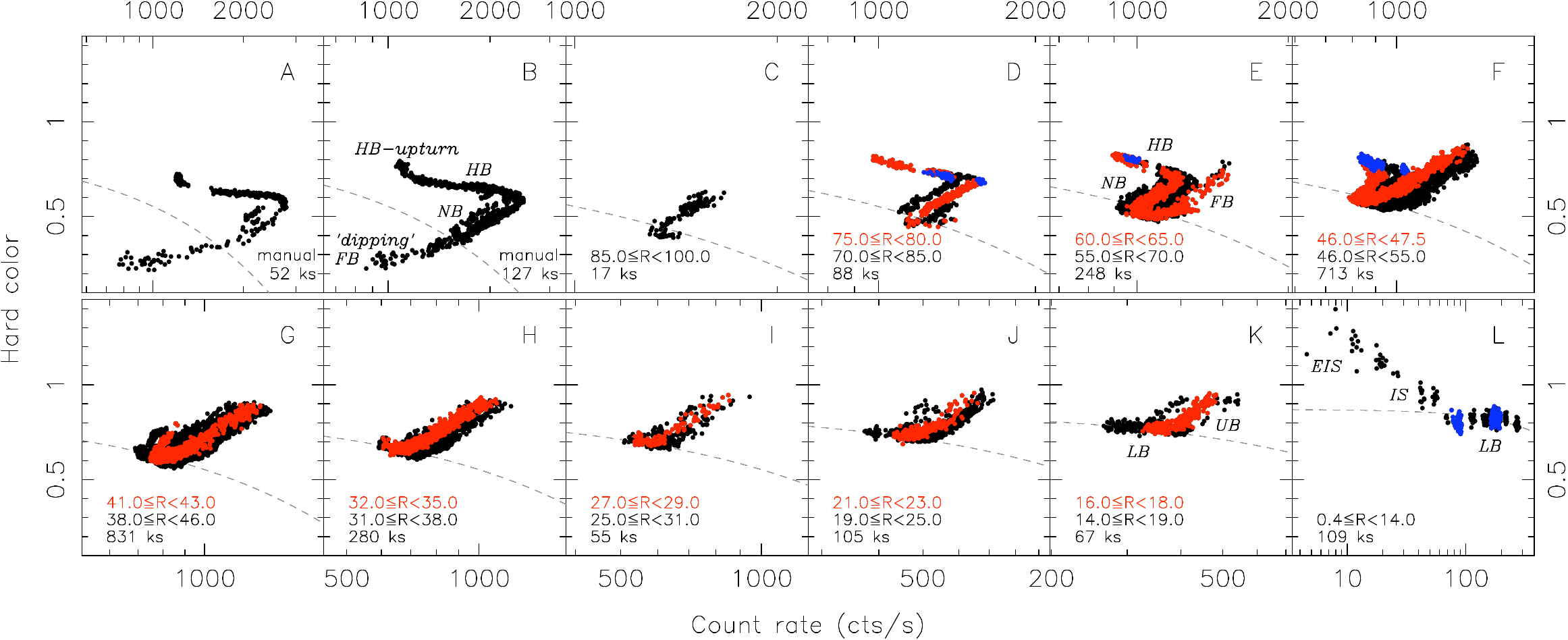}
\caption{The evolution of the HID tracks of \sj\ as a function of low-energy count rate.  Black data correspond to the same data as shown in Figure \ref{fig:cd}, whereas the red data represent narrower count rate intervals with reduced effects of secular motion. The selection criteria for the tracks are given at the bottom of each panel, together with the total amount of exposure time (only for the entire selection). Diagrams C-L are based on selections of the low-energy ($\sim$2.0--2.9 keV) count rate $R$; diagrams A and B are based on manual selections (before day 29).  For the intensity we used the total PCU2 count rate; for the definition of hard color see the caption of Figure \ref{fig:cd}.  Data points represent 64 s averages for high count rates, increasing to 1024 s averages for the lowest count rates. The blue points indicate observations in which kHz QPOs were detected. The dashed gray line indicates the approximate path followed by the normal/flaring branch vertex during the transition from Cyg-like Z to Sco-like Z and, finally, atoll source behavior.}
\label{fig:hid}
\end{figure*}

Twelve of the thirteen CDs and HIDs are shown in Figures \ref{fig:cd} and \ref{fig:hid}, respectively, in order of decreasing low-energy count rate. The average {\it total} count rate of the tracks decreases (monotonically) in the same direction as well. The twelve CDs and HIDs will be referred to as selections A-L throughout the remainder of the paper. Except for observations with low-energy count rates below 0.4 cts\,s$^{-1}$ (i.e., the thirteenth count rate interval), which have data with large errors bars, these selections represent our entire \xte\ data set of \sj. The count rate intervals used for each CD/HID are shown near the bottom of each panel, except for selections A and B, which were based on manual selections and had peak low-energy count rates of $\sim$100--165 cts\,s$^{-1}$. The total amount of time in each selection is shown as well.  The horizontal branch observations marked with red data points in the light curves in Figure \ref{fig:lc} are shown in red in Figure \ref{fig:cd} as well. Compared to the CDs shown in LRH09, which were based on broad time selections, the set of CDs shown in Figure \ref{fig:cd} reveals a more detailed evolution of the CD tracks.  We note that the ordering of the panels in Figures \ref{fig:cd} and \ref{fig:hid} does not reflect the evolution of the source as a function of time; as can be seen in Figure \ref{fig:lc}a, \sj\ moved back and forth between different selections (C--H) for most of the outburst. 

 While our low-energy count rate intervals result in relatively narrow CD tracks, there is still considerable secular motion within some of the HID tracks, making it harder to follow the evolution of these tracks than in the CDs. For this reason we created subset HIDs from narrower count rate selections within selections D--K, which are shown in red in Figure \ref{fig:hid}.

Significant evolution can be seen in the shape of the CD/HID tracks. It has already been reported that the CD/HID tracks of \sj\ evolved from Cyg-like Z tracks, to Sco-like Z tracks, and finally into atoll-like tracks \citep[H07,][LRH09]{howial2007}, which can clearly be seen in Figures \ref{fig:cd} and \ref{fig:hid}. Typical Z and atoll branches have been marked in a few of the panels of Figure \ref{fig:cd}. In the following we take a more in-depth look at the evolution of the CD/HID tracks and also compare the CD/HID tracks of \sj\ to those found in other NS-LMXBs.

\subsubsection{Comparison with CDs/HIDs of other NS-LMXBs}\label{sec:comparison}

To put the evolution of the CD/HID tracks of \sj\ into a broader context, we compare the various CD/HID tracks observed in \sj\ with those seen in other NS-LMXBs. For this comparison we mainly refer to work that is based on {\it RXTE}/PCA data, as comparison with data from other satellites is complicated by differences in energy bands and instrument responses. We also refer to the CDs/HIDs presented in Figure \ref{fig:overview}, which were taken from Fridriksson et al. (2010, in prep.).

\begin{itemize}

\item{Selections A \& B: The overall shapes of the CD/HID tracks in these two selections resemble those seen for Cyg X-2 in its high luminosity state \citep{wiva2001,murech2002,pisaka2002,lequso2008}. They also show similarities to the tracks observed in the Cyg-like Z sources GX 340+0 \citep{jovawi2000}, GX 5--1 \citep[][see also Fig.\ \ref{fig:overview}a]{jovaho2002}, however, with some notable differences in the properties of the flaring branch. The flaring branch in the CD of these two sources is usually oriented horizontally, rather than pointing downward like in \sj, although downward-pointing extensions to the flaring branch have been observed for these two sources (the so-called `dipping' flaring branch). Moreover, in GX 5--1 and GX 340+0 the count rate increases when they move onto the flaring branch (at least up to the `dipping' flaring branch), whereas the count rate decreases strongly on the flaring branch in selections A and B. Such dramatic drops in intensity on the flaring branch have only been seen in Cyg X-2 during its high luminosity state. We note that the track of selection A is incomplete, as evidenced by the visible gaps.}

\item{Selection C: This selection contains relatively little data; the track is likely to be incomplete and therefore difficult to compare with other sources. However, the properties of the flaring branch in the CD (short length and horizontal orientation) and HID (count rate increase) suggest that it might be part of a track more similar to those of GX 5--1 and GX 340+0 than the ones seen in selections A and B.}

\item{Selections D to H: These selections show CD tracks that are similar to those of the Sco-like Z sources GX 17+2 \citep[][see also Fig.\ \ref{fig:overview}b]{hovajo2002}, Sco X-1 \citep{vaswzh1996,brgefo2003}, and GX 349+2 \citep[][see also Fig.\ \ref{fig:overview}c]{onkuso2002,agsr2003}; they also show similarities to those of Cyg X-2 in its low luminosity state \citep{kuvava1996,murech2002} and  GX 13+1 \citep{murech2002,screva2003}. The latter is often classified as a bright atoll source, but also displays Z source characteristics \citep{howiru2004}. Note that in the HIDs of selections D and E the horizontal branch is still relatively long and close to being horizontal, similar to the Cyg-like Z sources, but unlike what is observed for the Sco-like Z sources. The properties of the horizontal branch become more Sco-like in the HID of selection F, where it turns upward and shortens. Large intensity increases associated with the flaring branch, which are typical for the Sco-like Z sources, only start to become visible in selections E and F.}

\item{Selection I: In this selection only hints of the normal branch can be seen in the CD and HID, in addition to a full-fledged flaring branch. The {\it EXOSAT} observations of GX 349+2 analyzed by \citet{hava1989} reveal a similarly shaped track. Overall, the track appears to be intermediate to the Z source flaring branch and the banana branches displayed by bright atoll sources such as GX 9+1 and GX 9+9 (see below).}

\item{Selections J \& K: No normal branch is seen in these selections and the curvature in the remaining branch in the CDs,  while still
retaining some characteristics of the flaring branch (particularly in selection J), is also reminiscent
of the lower and upper banana branches displayed by bright atoll sources such as GX 9+1 and GX 9+9 \citep[see also Fig.\ \ref{fig:overview}d]{hava1989} and to a lesser extent GX 3+1. Curvature of the track is stronger in selection K, where the lower banana branch becomes more prominent. The intensity shifts shown by the (upper) banana branch in the HIDs, which are the result of secular motion, are also seen for the bright atoll sources; in particular, the HID of selection K is similar to that of the atoll source 4U 1735--44, which
also shows secular motion of the upper banana branch (Fig.\ \ref{fig:overview}e).}

\item{Selection L: This selection shows a combination of branches (lower-banana, island state, and extreme island state) that is frequently seen in persistent atoll sources such as 4U 1636--53 and 4U 1705--44 \citep[see also Fig.\ \ref{fig:overview}f,g]{behomo2007,hokava2009} and transient atoll sources such as Aql X-1 and 4U 1608--52 \citep{lireho2007,gldogi2007}. For a better comparison with these sources we show the combined CDs and HIDs of selections J--L in Figure \ref{fig:flaring}. \citet{gldogi2007}
found considerable variation in the orientation of the island state among the atoll sources they studied. In \sj\ the island state is vertical in the CD and pointing to the upper left in the HID, similar to 4U 1735--44 (Fig.\ \ref{fig:overview}e) and 4U 0614+09 \citep{gldogi2007}.} 
\end{itemize}

\subsubsection{Evolution of tracks and branches}\label{sec:evolution}

The above comparison with other NS-LMXBs confirms our earlier reports that the CD/HID tracks of \sj\ evolved from Cyg-like Z to Sco-like Z, and finally became atoll-like. What follows is a brief overview of the most significant aspects in the evolution of the CD/HID tracks and branches.

\begin{figure}[t]
\vspace{0.2cm}
\epsscale{1.0}
\plotone{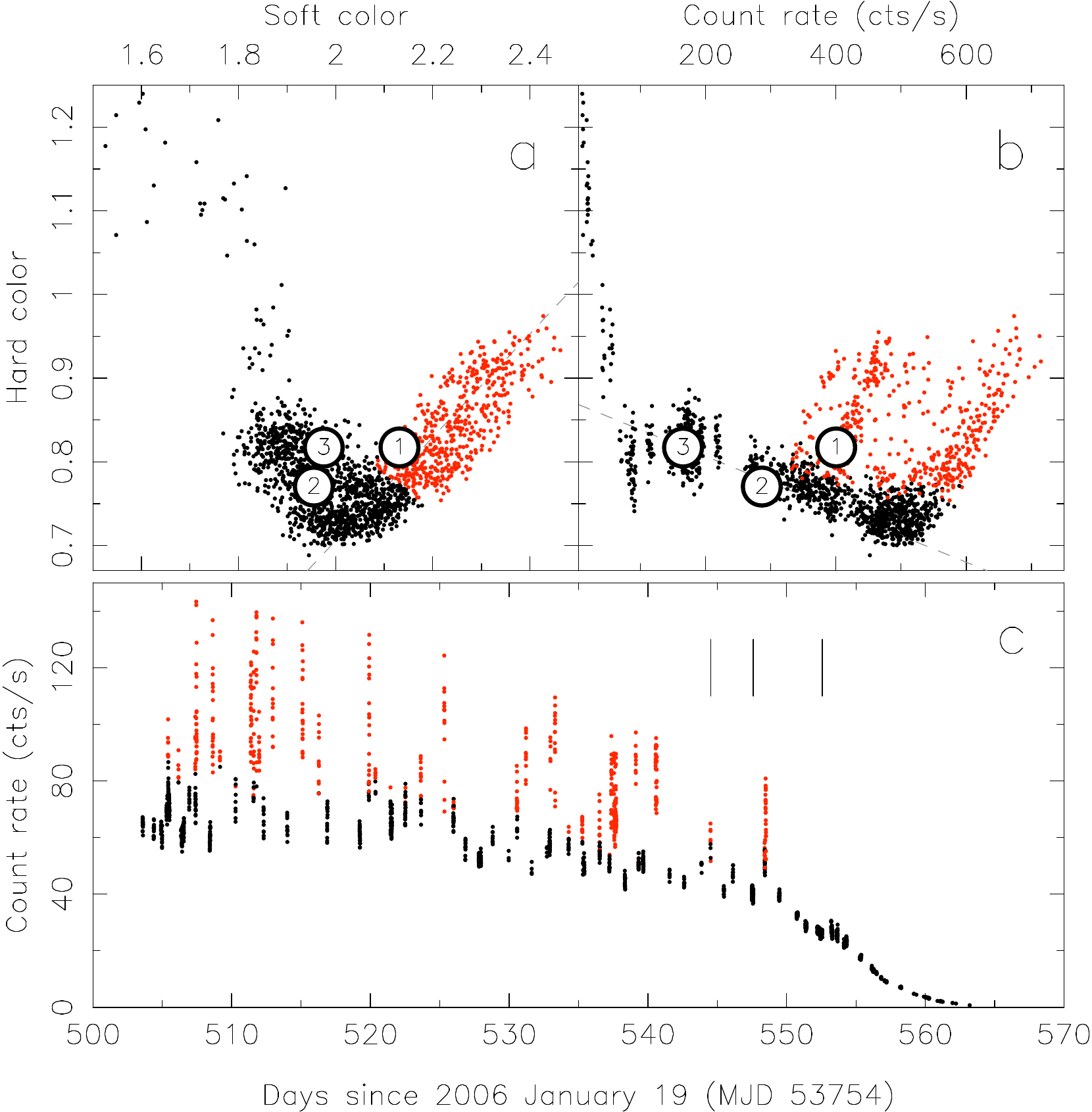}
\caption{The combined CD (a), HID (b), and 9.4--18.1 keV light curve (c) of  selections J--L The red-colored data points show that the part of the atoll source track that has historically been classified as the ``upper banana"  corresponds to the last episodes of flaring in the Z source phase. This flaring abruptly subsides below total count rates of $\sim$300 cts\,s$^{-1}$ (b), corresponding to a high energy count rate of $\sim$40 cts\,s$^{-1}$ (c). The numbered circles in the CD and HID show the location of the three type I X-ray bursts that were detected. The times of the three bursts are indicated by the three vertical lines in the light curve; the bursts themselves are not shown.}
\label{fig:flaring}
\end{figure}

The horizontal branch in the high-intensity and Cyg-like selections A and B shows large intensity swings that only results in minor color changes. Selections A and B are also the only two selections to show an extension to the horizontal branch that previously has been referred to as the horizontal branch `upturn' (H07, LRH09). In the CD this upturn appears to be more of a separate branch than in the HID, similar to what is seen in GX 5-1 and in the {\it EXOSAT} data of Cyg X-2 \citep{kuvava1996}.  The horizontal branch is not observed in selection C, probably due to the short total exposure time. In the CDs of selections D-F the orientation of the horizontal branch is rotated in a clockwise direction with respect to selections A and B, while the intensity swings have become less pronounced.  The upturn is no longer present in these selections, and the horizontal branch itself is no longer visible in selection G. 

The normal branch gradually becomes shorter in the CDs and HIDs from selection A to H, while it rotates slightly in a counter-clockwise direction. As in the case of the horizontal branch, the shortening of the normal branch in the HID means that the intensity swings become less pronounced. The normal branch is only marginally detected in selection I. As was already pointed out by H07,  the vertex between the normal and flaring branches moved along a more or less straight line in the CD during the first part of the outburst, as the result of the secular changes. As can be seen in Figure \ref{fig:cd}, where we have represented the path of the vertex with a dashed line, this also appears to be true for the remainder of the Z source phase of the outburst. LRH09 showed that this vertex also followed a straight line in their HID. With our color definition (i.e., chosen energy bands) the path followed by the vertex in our HID shows some deviations from a straight line (especially in the three highest count rate selections), but it is still rather close. The dashed line in Figure \ref{fig:hid} represents the approximate path of the vertex; it is a straight line, but it appears curved because of the logarithmic scale on the horizontal axis.

The flaring branch undergoes large changes in both the CD and HID and is crucial in linking Z and atoll source behavior. In selections A and B motion onto the flaring branch corresponds to drops in the intensity, and in the HID the flaring branch appears to be an extension of the normal branch. In the CD of selection A the flaring branch points downward, but it rotates in a counter-clockwise direction in the lower count rate selections. Starting in selection C, motion onto the flaring branch corresponds to count rate increases, which become more pronounced in selections E and F. This flaring eventually starts to become weaker (see also the high-energy light curve in Figure \ref{fig:lc}b), but it remains present through selection K.   While selections J and K can be classified as atoll-like (\S\ref{sec:comparison}), it is important to note that the flaring that has long been regarded as being characteristic of the Sco-like Z sources is still present in these lower count rate intervals. In the framework of atoll sources, the branches traced out as a result of this flaring would be referred to as the atoll upper-banana branch, but as Figures \ref{fig:lc}, \ref{fig:cd}, and \ref{fig:hid} show, there is no sharp transition between Z and atoll behavior. To better describe the rapid evolution during the last phase of the outburst and emphasize the relation between the Z source flaring and the atoll upper-banana branch, we show the combined CD, HID, and light curve of selections J, K, and L in Figure \ref{fig:flaring}, resulting in a `complete' atoll track. Points on the upper-banana branch of the atoll in the CD were colored red; from the HID and light curve it can be seen that these upper-banana branch data points correspond to the last few flares that were observed in \sj\ (compare also with Figs.\ \ref{fig:lc}b and \ref{fig:hid}). This is the first time that the Z source flaring branch has been observed to evolve into the atoll source upper banana branch and it confirms earlier suspicions \citep{hava1989} that the two branches might be one and the same phenomenon.

The lower banana branch starts to form around the time of the last flares in selections J and K.   As the total count rate drops by a factor of $\sim$3 from selection K to selection L, the source moves away from the diagonal line followed by (the secular motion of) the normal/flaring branch vertex in selections A to I (dashed lines in Fig.\ \ref{fig:cd}). However, in the HID the lower banana branch continues along the path traced out by the secular motion of this vertex (dashed lines in Fig.\ \ref{fig:hid}), as already pointed out by LRH09. The transition to the atoll island state occurs in selection L during a rapid decrease in count rate, and is marked by a sudden increase in hard color. It is followed by the extreme-island state, in which the hard color becomes more or less constant (although poorly constrained because of the low count rates).

\begin{figure}[t]
\vspace{0.5cm}
\centerline{\hbox{
\includegraphics[height=4.5cm]{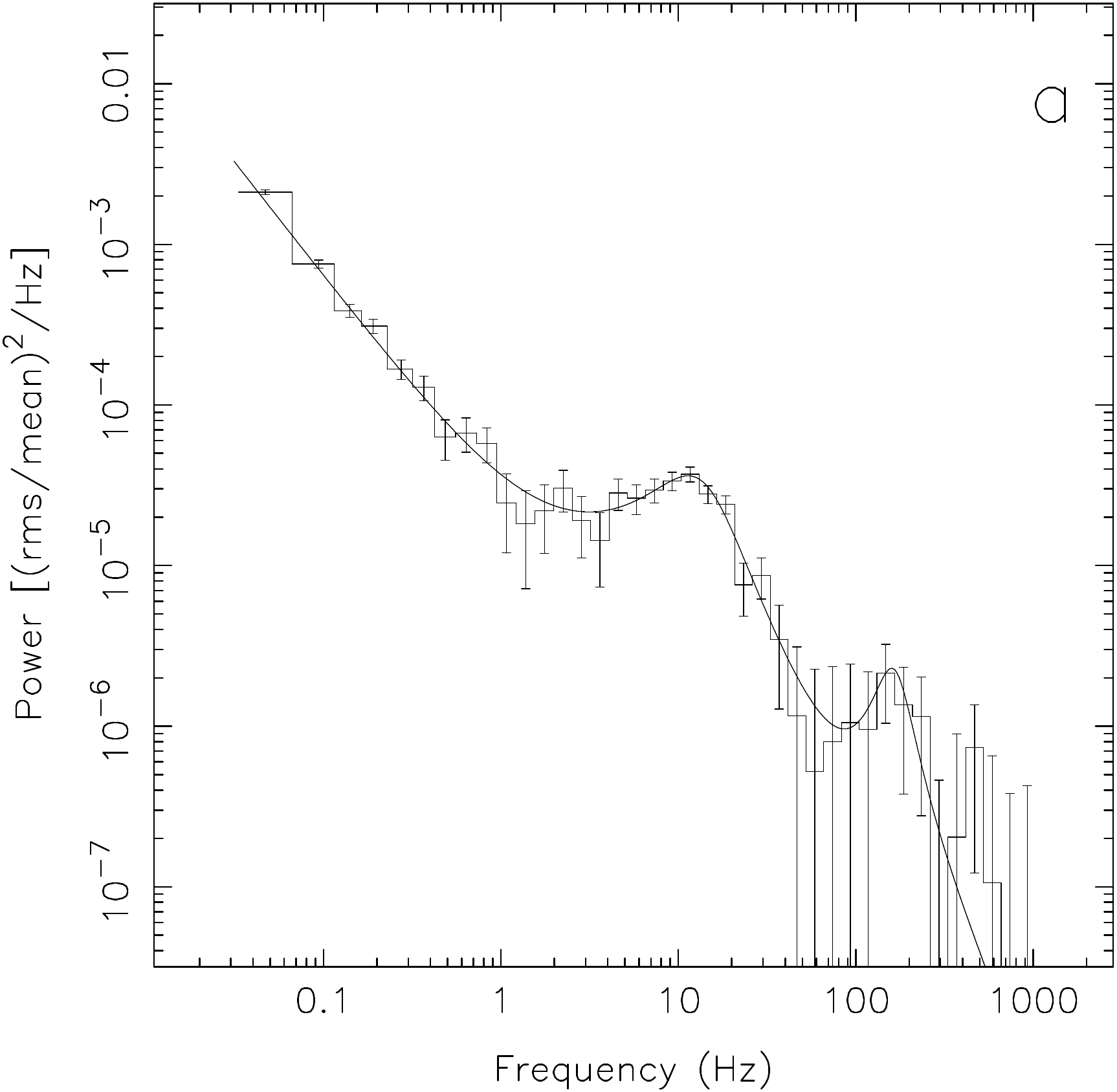}
\includegraphics[height=4.5cm]{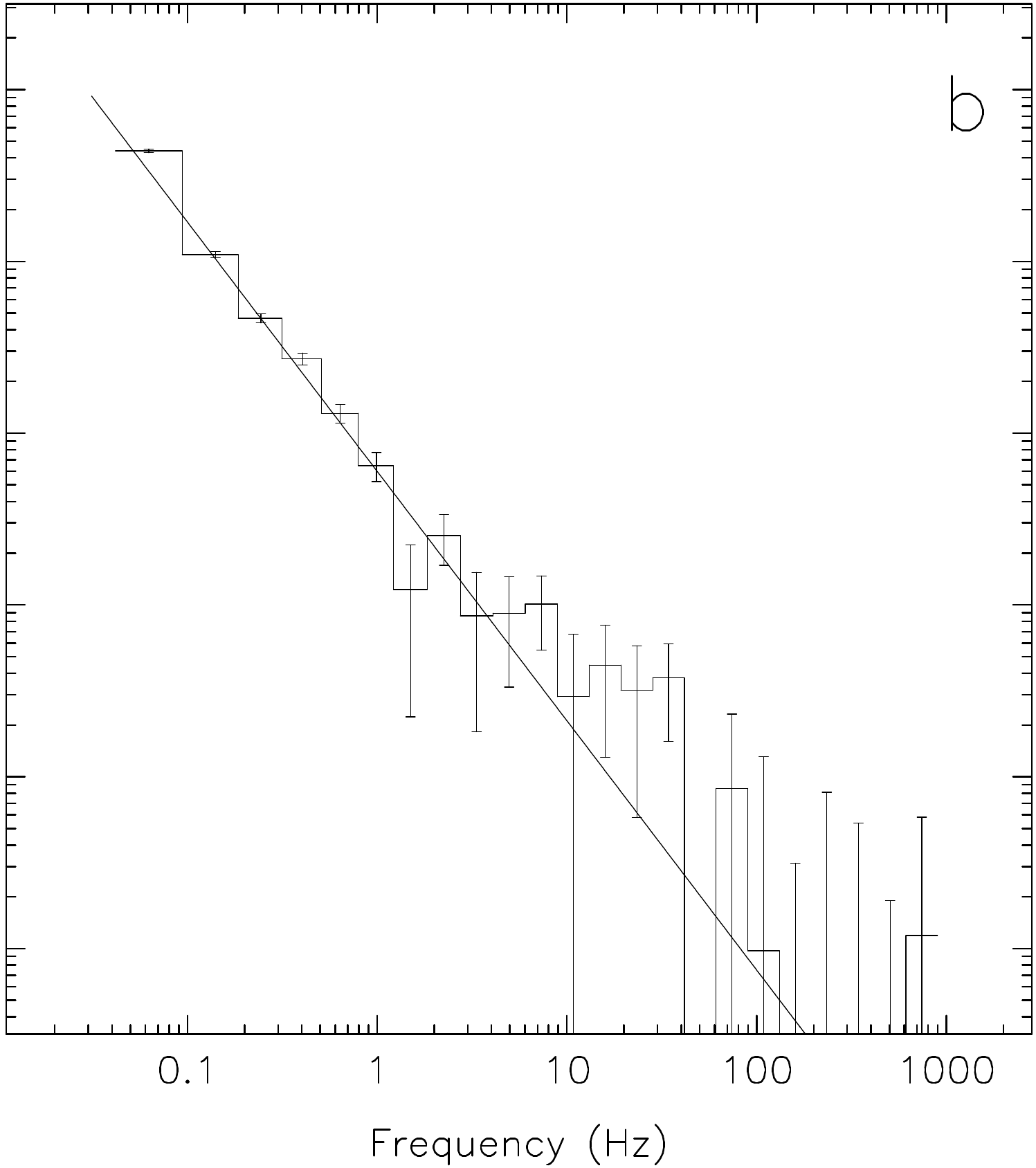}
}}
\caption{Power spectra from the lower-banana (a) and upper-banana branch (right) of the GX-atoll-like selections I and J, shown with their best fits. The feature around 160 Hz in panel (a) is somewhat suggestive of a ``hectohertz QPO" \citep{vavadi2002}, but it is only detected at a 1.8 $\sigma$ level.  }
\label{fig:low-freq}
\end{figure}

\subsection{Summary of CD/HID evolution}

Figures \ref{fig:cd} and \ref{fig:hid} show that \sj\ is the first NS-LMXB in which we can study the entire evolution from Z source to atoll source behavior in a single source. Cyg-like Z source behavior, characterized by large intensity swings on the horizontal and normal branches, the presence of a dipping flaring branch, and an upturn from the horizontal branch, is observed at the highest count rates. The transition to Sco-like Z source behavior is rapid ($\sim$5 days) with maximum total count rates dropping by a factor of $\sim$1.5; this transition mostly involves changes in the orientation of the branches in the CD and HID, as well as the disappearance of the horizontal branch upturn. The transition from Sco-like Z behavior to atoll source behavior is much slower ($\sim$months). Figure \ref{fig:flaring} shows that the last instances of the Z source flaring branch, at low overall count rates, can also be classified as the atoll source upper banana branch, providing an important link between the two subclasses. The Sco-like Z to atoll transition is marked by the disappearance of the horizontal branch, normal branch, and flaring/upper-banana branch, as the maximum total count rates dropped by a factor of $\sim$3. In terms of the selections A-L this is a systematic and
monotonic process; of course, during its initial rise and in subsequent decay the source moves back and
forth between some of these selections. The time between the last detection of the flaring branch/upper banana branch to the transition from the lower-banana branch to the island state (and later the extreme island state) is $\sim$7 days, with the count rate dropping by another factor of $\sim$3. The evolution during the atoll phase presents an important departure from the evolution seen in the Z phase of the outburst. While in the Sco-like Z source phase the change in low-energy count rate results in gradual changes in the CD/HID tracks (except for occasional visits to the horizontal branch), in the atoll phase (after the flaring subsides) the low-energy count rate changes result in motion {\it along} the CD/HID track. In the HID, motion along the lower-banana branch even appears to be a continuation of the gradual motion of the Z source normal/flaring branch vertex (this is not the case in the CD, however).

\subsection{Low-frequency variability}\label{sec:low}

\begin{figure}[t]
\vspace{0.5cm}
\epsscale{0.8}
\plotone{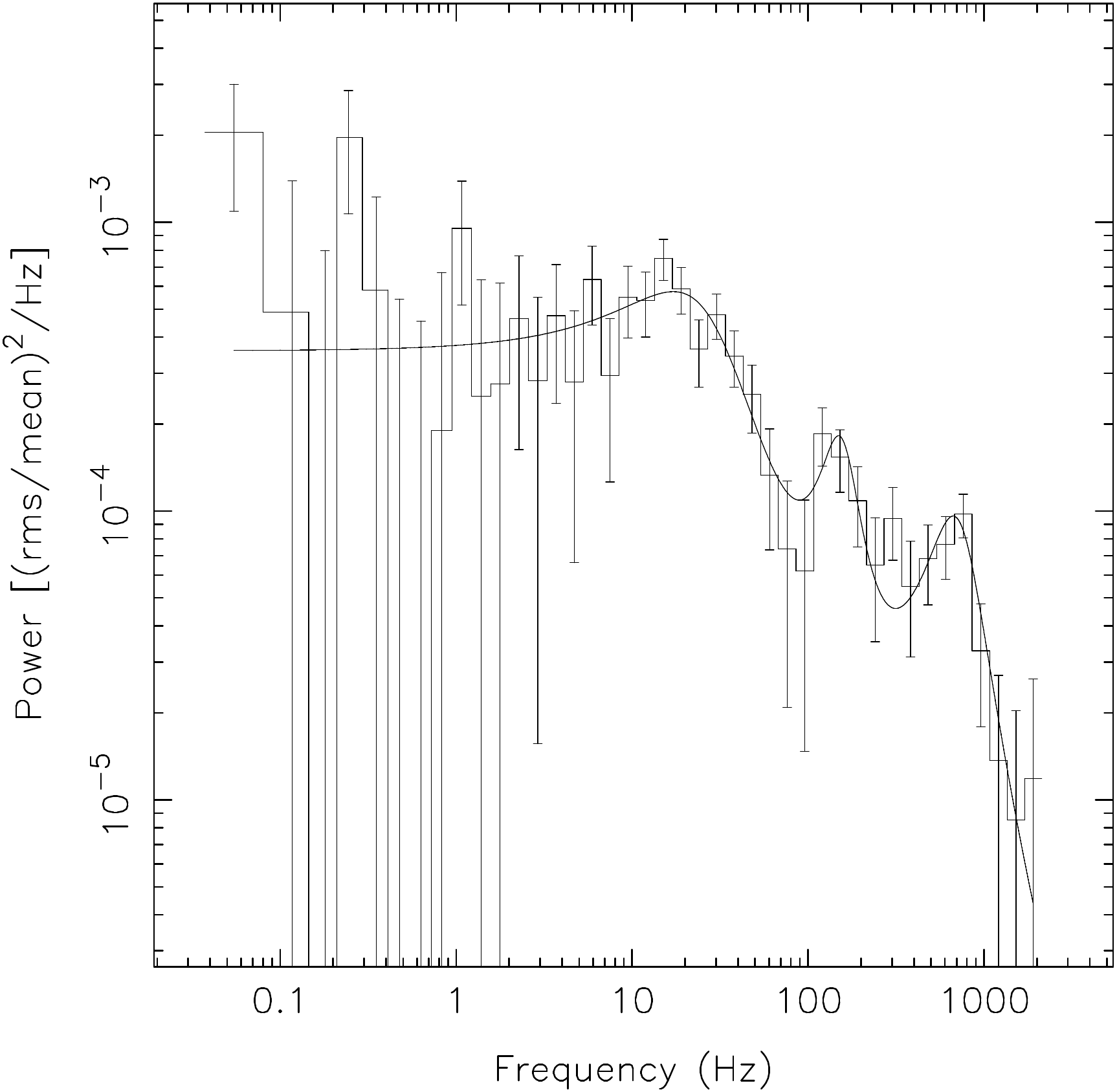}
\caption{The island state power spectrum of atoll-like selection L. The solid line shows the best fit (consisting of three Lorentzians) to the power spectrum.}
\label{fig:island}
\end{figure}

A low-frequency variability study (in the 6.9--60 keV band) of the Cyg-like tracks and some of the Sco-like tracks was already presented in H07. Two types of low-frequency QPOs were observed in those tracks, the horizontal branch oscillations ($\sim$10--60 Hz) and the normal branch oscillations ($\sim$7--9 Hz). These QPOs appeared on the appropriate branches, confirming our branch identification and our interpretation of the highest count rate tracks as Z-like tracks. We did not perform a similarly detailed analysis for the remainder of the outburst, but a visual inspection of the power spectra indicates that both types of low-frequency QPOs disappear as the overall count rate decreases. The normal branch oscillations disappear within selection E, where they are seen in the high count rate HID tracks, but not the lower count rate ones. Indications of horizontal branch oscillations are still seen in selections F and G, but not in the lower count rate selections H--L.

In Figure \ref{fig:low-freq} we show examples of 2--60 keV lower-banana (a) and upper-banana branch (b) power spectra from the combined bright-atoll-like selections I and J. The shapes are consistent with the power spectra of other atolls in these spectral states \citep{va2006}. The lower-banana power spectrum shows a broad  (Q=0.7$\pm$0.1) peak at 13.9$\pm$0.7 Hz, on top of a power-law shaped ($\propto \nu^{-1.41\pm0.04}$) noise continuum. The overall variability is weak, with an integrated power rms (1/16--100 Hz) of $\sim$3\% (2--60 keV). The upper-banana power spectra could be fitted with a single power law ($\propto \nu^{-1.48\pm0.03}$) and the 1/16--100 Hz rms was $\sim$2.4\%. 

A similar study for the atoll-like selections K and L suffers from low count rates, and high quality broadband power spectra are difficult to produce. The shape of the power spectrum of the lower-banana branch of selection L (from observations with total count rate above 65 counts/s/PCU) is consistent with that of the lower-banana branch of selections I and J; we measure an rms amplitude of 5.0$\pm$0.3\%, which is also consistent with that seen in atoll sources \citep{va2006}. The island state power spectrum was averaged from observations with hard color 0.9--1.1 and total count rate above 35 counts/s/PCU. The resulting power spectrum is shown in Figure \ref{fig:island}. It could be fit with a model consisting of three broad ($0<Q<1.8$) Lorentzians, with frequencies of 27$\pm$4, 155$\pm$19, and 720$\pm$50 Hz. The 720 Hz feature is probably a (broadened) lower or upper kHz QPO (see below), whereas the 155 Hz feature could be a hecto-hertz QPO, a component that is frequently seen in this state \citep{vavadi2002,vavame2003}. The broad 27 Hz feature likely corresponds to a low-frequency band-limited noise component. We measure a 1/16--100 Hz rms amplitude (2--60 keV) of 16.3$\pm$0.7\%, which combined with the overall shape of the power spectrum is consistent with the island state of other atoll sources. For the extreme island state we averaged power spectra from observations with total count rates between 4 and 19 counts/s/PCU (with hard color $\sim$1.1--1.3). We measure an rms amplitude of 29$\pm$2\%, also similar to what is found in the extreme island state of other atoll sources. For reference, the highest rms for the Z-like selections was measured on the horizontal branch upturn in selections A and B: 15.0$\pm$0.2\%.

\subsection{Kilohertz QPOs}\label{sec:khz}

\begin{figure}[t]
\vspace{0.2cm}
\epsscale{1.1}
\plotone{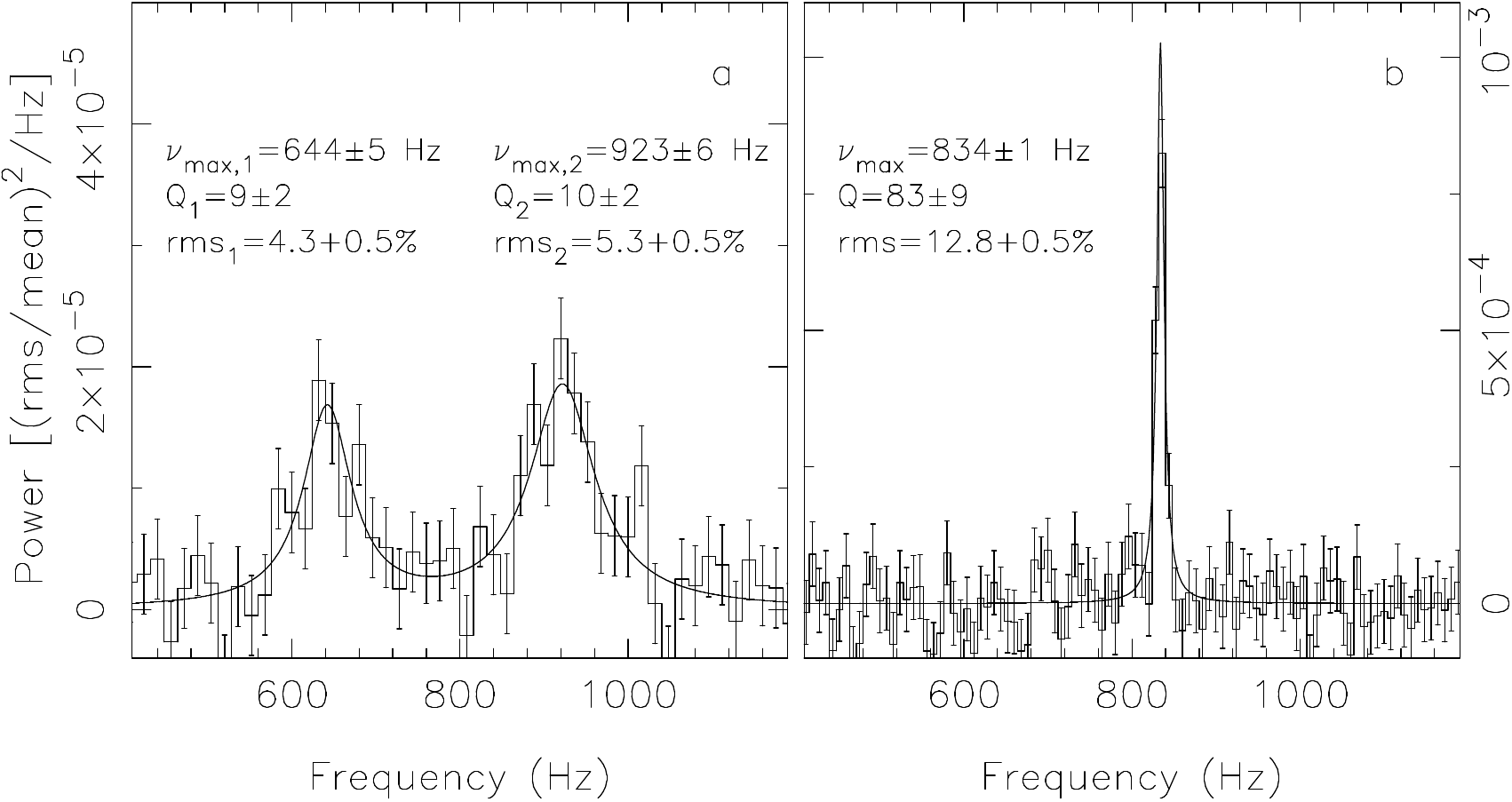}
\caption{Two examples of kHz QPO detections in the 6.9--60 keV band. On the left we show a pair of kHz QPOs from the horizontal branch of selection F (observations 92405-01-40-04 and 92405-01-40-05 combined). On the right we show a single kHz QPO from the lower banana branch of selection L (observation 93703-01-02-05). Fit parameters for all QPOs are shown. Notice the difference in vertical scale between the two panels.}
\label{fig:khz}
\end{figure}

Following H07, the 2--60 keV and 6.9--60 keV power spectra of all individual observations were visually inspected for high-frequency (kHz) QPOs. In addition to the (marginal) detections reported in H07, we find two pairs of kHz QPOs and six single kHz QPOs. The observation IDs of all ten $\gtrsim3\sigma$ kHz QPO detections (pairs and singles) are listed in Table \ref{tab:khz}, together with their CD/HID selection and QPO frequencies; frequencies are from fits to the power spectra of the 6.9-60 keV energy band. The locations  of these observations in the CDs/HIDs are shown in blue in Figures \ref{fig:cd} and \ref{fig:hid}. One new pair was detected on the horizontal branch of selection D and the other was detected in two consecutive observations (made about a day apart) on the horizontal branch of the Sco-like track of selection F; the power spectrum of the second pair is shown in Figure \ref{fig:khz}a. The QPO properties are consistent with those of the pairs reported in H07, which were detected on the horizontal branch or close to the horizontal/normal branch vertex of the Sco-like CD/HID tracks of selections D, E, and F. The kHz QPO pairs have peak separations between  274$\pm$11 Hz and 319$\pm$7 Hz, and the average peak separation is $\sim$285 Hz, consistent with what is found in other NS-LMXBs \citep{mebe2007}. The six new single kHz QPO detections were all made 
on the lower-banana of the atoll-like CD/HID track of selection L. The 720 Hz feature in the combined island state power spectrum (\S\ref{sec:low}) was not detected significantly in individual observations. Besides the fact that these single peaks were detected in an entirely different part of the CD, their properties also differ considerably from the peaks detected earlier in the Sco-like CD tracks. As can be seen from the representative example in Figure \ref{fig:khz}b, these peaks have rms amplitudes that are two to three times higher than the peaks in the Z like tracks and also have much higher Q-values, consistent with the (lower) kHz QPOs found in atoll sources \citep{me2006,va2006}. Their frequencies ranged from $\sim$650 Hz to $\sim$850 Hz, and in dynamical power spectra of two individual observations (with lengths of $\sim$8-15 ks) the frequency was found to drift by up to $\sim$80 Hz. \citet{sameal2010}, combining all kHz QPO data from the lower banana branch and employing a shift-and-add technique to align the drifting peaks \citep{mevava1998a}, report the marginal detection of a second peak, with a peak separation on 258$\pm$13 Hz. Finally, we note that averaging power spectra based on selections in the CD/HID  \citep[e.g.,][]{hovajo2002} revealed additional indications for kHz QPOs on the horizontal and normal branches of the Sco-like selections D--F; however, a full study of this kind is beyond the scope of this paper.

\section{Discussion}\label{sec:discussion}

\begin{table}[t]
\caption{Kilohertz QPO detections in \sj}
\begin{center}
\begin{tabular}{lcc}
\hline
Observation ID & Selection & $\nu$ (Hz)  \\
\hline
\hline
91442-01-07-09 & D & 641$\pm$5, 915$\pm$10 \\
92405-01-01-02 & F & 613$\pm$4, 932$\pm$6 \\
92405-01-01-04 & E & 761$\pm$6  \\
92405-01-02-03 & D & 622$\pm$9, 911$\pm$12 \\
92405-01-03-05 & F & 617$\pm$11,   908$\pm$9 \\
92405-01-40-04$^a$ & F & 644$\pm$5, 923$\pm$6\\
93703-01-02-04 & L & 852$\pm$2  \\
93703-01-02-11$^b$ & L &  $\sim$805  \\
93703-01-02-05 & L & 834$\pm$1  \\
93703-01-02-08 & L &  837$\pm$2  \\
93703-01-03-00$^c$ & L & $\sim$720 \\
93703-01-03-02 & L & 650$\pm$8 \\
\hline
\end{tabular}
\end{center}
$^a$ Fitted together with power spectrum of observation 92405-01-40-05\\
$^b$ QPO moved by $\sim$60 Hz during observation\\
$^c$ QPO moved by $\sim$80 Hz during observation
\label{tab:khz}
\end{table}

\subsection{Identification of Z and atoll phases}\label{sec:identification}

In \S\ref{sec:results} we presented an overview of the 2006--2007 outburst of the transient NS-LMXB \sj. By comparing the CD/HID tracks of \sj\ with other sources, we showed that the source made a rapid transition from Cyg-like to Sco-like Z source tracks, followed by a slower transition to various types of atoll source tracks. The transitions between Z
and atoll and further subclass divisions are not monotonic but closely
track the 2--2.9 keV count rate; we note that the spectral fits by LRH09  suggest that for most branches the count rate in this energy band is dominated by the contribution from the accretion disk component. During the outburst, the count 2--2.9 keV rate initially goes down rapidly, which
is linked to the rapid Cyg-like to Sco-like transition; later it goes up
and down superimposed on the gradual decay, and this is associated with
the source moving back and forth between subclasses. We interpret the
2--2.9 keV count rate as a proxy for the mass accretion rate component \mdot,
possibly that from outer disk to the inner parts of the accretion flow, that determines subclass. \sj\ is the first NS-LMXB in which such transitions between subclasses in the CD/HID have been followed in detail (see also H07, LRH09). 
The properties of the low- and high-frequency variability observed in \sj\ (\S \ref{sec:low} and \ref{sec:khz}; see also H07 and LRH09) broadly support our identification of the various Z-like and atoll-like tracks, although there are a few minor differences; we expected to see kHz QPOs in the Cyg-like Z tracks (see \S \ref{sec:disc-khz}) and normal branch oscillations on the normal branch of selection F, but none were detected. 

Three type I X-ray bursts were detected in the entire \xte\ dataset of \sj\ \citep{lialho2009}. We have marked the times of the three bursts in the light curves in Figure \ref{fig:lc} and also marked their location in the CD and HID in Figure \ref{fig:flaring}. As can be seen, the bursts occurred late in the outburst, with two bursts occurring on the lower  banana branch and one occurring on the atoll upper banana or Z source flaring branch (see \S \ref{sec:evolution}).
The fact that the only three detections of type I X-ray bursts were made in the atoll phase of the outburst is in line with the fact that atoll sources are more prolific bursters than Z sources \citep{gamuha2008}, and further supports our identification of the final phase of the outburst as atoll-like.  \citet{lialho2009} also derive a distance estimate of 8.8$\pm$1.3 kpc from two radius expansion bursts (the latter two of the three), implying that the persistent luminosity just before the bursts was $\sim$10\% \ledd, also consistent with atoll sources. 

A spectral analysis of the complete \xte\ data set of \sj\ is presented in LRH09. They find that near the end of the outburst, below luminosities of $\sim$15\% \ledd, the spectral behavior of \sj\ was similar to that of the atoll-type transients Aql X-1 and 4U 1608--52 \citep{{lireho2007}}. Notably, both the disk and boundary layer components closely follow $L\propto T^4$ on the atoll lower banana branch, implying a relatively constant emission area for both components as \mdot\ changes. The spectral results of the Z source stage could not be compared to other Z sources, as the spectra of the latter have not yet been fitted with the model used by LRH09.

\subsection{The nature of NS-LMXB subclasses}\label{sec:nature}

For each of the CD/HID tracks observed in \sj\ (Figs.\ \ref{fig:cd}, \ref{fig:hid}, and \ref{fig:flaring}) similarly shaped tracks can be found among the persistent NS-LMXBs (\S\ref{sec:comparison}).  Moreover, except for sources that suffer from high internal absorption (i.e., dipping and/or eclipsing sources), we are not aware of sources that show CD/HID tracks that are wildly different from the ones seen in \sj. This seems to suggest that the evolution of the tracks seen in \sj\ is representative of the entire class of NS-LMXBs, which we will further explore in an upcoming paper (Fridriksson et al.\ 2010, in prep.).

The gradual transition from Sco-like Z to atoll behavior observed in \sj\ argues against a sharply defined boundary between these two NS-LMXB subclasses. In particular, the flaring that is apparent in the high-energy light curves of \sj\ persists across the Z/atoll periods of the outburst, as it is responsible for both the Z source flaring branch {\it and} the atoll source upper banana branch. This flaring rapidly becomes weaker in the atoll phase, where it eventually subsides around a luminosity of $\sim$0.15\,\ledd\ (\S\ref{sec:evolution}; luminosity taken from LRH09). In addition to the disappearance of the flaring, there are two additional points in the evolution of \sj\ at which significant changes occur; one is the transition from Cyg-like to Sco-like Z behavior around \ledd, and the other is the switch between thermal- and non-thermal-dominated spectra (lower banana/island state transition) around 0.02\,\ledd\ (LRH09). While these points could serve as `natural' boundaries between different types of NS-LMXB subclasses, there are a number of persistent NS-LMXBs that regularly cross one or more of these boundaries, so their usefulness for (re-)defining NS-LMXB subclasses is therefore somewhat limited. 

As already  mentioned in \S\ref{sec:intro}, the disparity in the properties of the Z and atoll subclasses has been a long-standing problem: are differences in mass accretion rate alone sufficient to explain the differences in the phenomenology of the various subclasses, or are additional parameters needed? Differences in the neutron-star magnetic field have been suggested to be (partially) responsible for the differences between Z and atoll sources \citep{hava1989} and between Cyg-like and Sco-like Z sources \citep{pslami1995}. Diamagnetic screening by accreted matter  \citep{cuzwbi2001} appears  to be the only feasible way to affect (i.e. reduce) the external magnetic field of a neutron star on a short time scale. It cannot explain the existence of two classes of Z sources in terms of different magnetic field strengths, since fields in these sources should be completely buried as the result of their near-Eddington accretion rates. Explaining the Z to atoll transition is problematic as well. \citet{hava1989} suggested that atoll sources have a higher magnetic field than Z sources and to explain the observed Z to atoll transition, the underlying magnetic field needs to emerge again. However, the time scale on which this is expected to happen is much longer (100--1000 yr) than the observed time scale for the transition from Z to atoll in \sj. We therefore conclude that magnetic field strength is probably not responsible for the NSXBs subclasses.

Differences in viewing angle have been proposed to explain the existence of two types of Z sources, Cyg-like and Sco-like \citep[][and references therein]{kuvaoo1994,kuva1995}. Changes in the binary inclination are not expected on the time scale of weeks to months, so these can be ruled out as a source for the observed transitions between subclasses in \sj. However, a tilted precessing disk, such as thought to be present in, e.g., Cyg X-2 \citep{vrswke1988}, can lead to changes in the viewing geometry on a time scale of weeks to months. Combined with the effects of anisotropies and/or varying amounts of obscuration this could lead to changes in the morphology of CD/HID tracks and branches. However, such a model would require a complex evolution of the viewing angle to account for the observed sequence (in time) of the various sub-classes. Moreover, it cannot account for the systematic disappearance of the various Z source branches as is observed during the transition from Z to atoll. We conclude that viewing angle is probably also not responsible for the NSXB subclasses. We note that parameters such as neutron star mass and spin frequency should not change substantially during an outburst, as the amount of matter that is accreted is negligible in comparison to the neutron star mass.

Given the above, and taking also into consideration the transient nature of \sj, we conclude that transitions between the NS-LMXB subclasses in \sj\ are probably entirely driven by changes in \mdot. Although we cannot completely rule out that the persistent Z and atoll sources differ in, e.g., neutron star mass/spin/magnetic field, or viewing angle, our results indicate that these parameters are relatively unimportant to the phenomenology of \sj, and by extension, to that of the Z and atoll sources. Such parameters may, however, explain some of the differences between tracks in the CD/HIDs of sources with similar \mdot.

The discussion of the role of the various possible inflow/outflow components of the
accretion process in determining NS-LMXB subclasses and motion
along the CD/HID tracks is of course hampered by the lack of reliable
independent measures of the relevant inflow/outflow rates, so it is
difficult to draw firm conclusions, even from the rich phenomenology
observed in \sj. However, as we discussed above, for most of
the outburst of \sj\ the low-energy count rate appears to be a
fairly good tracer of the \mdot\ that determines the changes in the CD/HID
tracks.  It is plausible that this \mdot\ scales monotonically with some
overall representative average PCU2 ($\sim$2--60 keV) count rate level of the
CD/HID tracks  and their average bolometric luminosities (LRH09).
Under certain assumptions this would imply that Cyg-like Z source
behavior corresponds to the highest overall energy release and perhaps
\mdot\ values, followed by Sco-like Z source behavior and
then atoll source behavior. The above also implies that the horizontal
branch and its upturn are phenomena related to the highest \mdot\ values, although, as we discuss below, this does not necessarily mean
that along a single Z track these parts of the track represent the
highest values of the accretion rate relevant to motion along the Z.

\subsection{Spectral evolution along Z and atoll tracks}

The above interpretation in which an \mdot\ level determines the type of CD/HID track of NS LMXBs has implications for proposed models for the role of mass flow rates in motion (i.e. spectral evolution) along the Z tracks. Such a picture is obviously problematic for models in which the same \mdot\ level that determines the type of CD/HID also directly determines the position along the Z, with this \mdot\ either increasing monotonically from the horizontal to the flaring branch \citep{havaeb1990,vrraga1990}, as is still often assumed, or with \mdot\ changing in the opposite direction, as was recently proposed by \citet{chhaba2006} and \citet{jachba2009}. However, we cannot rule out that the accretion rate through some part of the inner accretion flow changes monotonically along the Z.

A suggested solution in which changes in a mass flow rate can explain both the changes in the type of the CD/HID tracks as well as motion along the Z track was discussed in H07. Their solution was based on a model by \citet{va2001} for the so-called `parallel tracks' phenomenon in NS-LMXBs \citep[see, e.g.,][for further information on this]{mevava1998a}, which assumes that there exists both a prompt and a filtered response to changes in the mass accretion rate \mdot$_d$ through the inner edge of the accretion disk. For a detailed discussion of the filtered response scenario and how it relates to motion along the Z we refer to H07, but we point out
an opportunity and a problem here.  The model predicts that branches
would be preferentially suppressed if there is a prolonged monotonic
decay in \mdisk, which is somewhat reminiscent of  the systematic
disappearance of the Z source horizontal and normal
branches.  However, the observed long term modulation in 2--2.9 keV flux
is not monotonic, and if it can be identified with \mdisk\ it should
alternatingly suppress the horizontal and the flaring branch, which is
not observed.

An alternative scenario, in which motion along the Z track is not governed by \mdot, was suggested by \citet{hovajo2002}. In such a scenario the parameter(s) responsible for spectral and variability properties, e.g., the inner disk radius, varies (vary) independently from \mdot. The spectral fits by LRH09 suggest that motion along the horizontal branch and the Sco-like flaring branch is consistent with occurring at a nearly constant \mdot, although this is not as clear for the normal branch and Cyg-like flaring branch. In any case, the spectral fits suggest that \mdot\ variations along individual Z tracks are very small. Based on this, LRH09 suggest that the different Z branches may be the result of various instabilities in the accretion flow, which could be related to an increase in the radiation pressure in different parts of the flow (inner disk, boundary layer). This scenario avoids the difficulties of \mdot\ having to explain both gradual changes in the
shape of the Z tracks and motion along the Z track at the same time. The observations of \sj\ require that the occurrence of the assumed (and unknown) instabilities themselves depends on the \mdot\ level (i.e., the instabilities resulting in the horizontal and normal branches only being present at the highest \mdot), but the proposed scenario provides no
explanation for the systematic suppression of horizontal and normal
branches as 2--2.9 keV count rate drops.

The above discussion has mostly focused on the role of \mdot\ in the Z source tracks. As already mentioned in \S\ref{sec:intro}, motion along atoll tracks appears to be mostly the result of changes in \mdot. The spectral fits by LRH09 support this. In this respect, it is interesting to point to the atoll lower-banana branch in the HID of \sj, which appears to be a continuation of the presumably \mdot\ driven secular motion observed in the Z-like phase of the outburst. However, the presence of hysteresis in the state transitions \citep{gldogi2007} and the phenomenon of parallel tracks \citep{mevava1998a} observed in other atoll NS-LMXBs suggest that a time averaged response to changes in \mdot\ might be involved as well \citep{va2001}.

\subsection{Kilohertz QPOs}\label{sec:disc-khz}

The luminosity range over which kHz QPOs are detected in \sj\ spans a factor of $\sim$15--20, which is wider than seen in any other NS-LMXB \citep{fovame2000}. However, there is a fairly large gap in luminosity in which no kHz QPOs are detected, corresponding to selections G--K. In selections G and H the source shows Sco-like Z behavior, but is missing the horizontal and upper normal branches, where most of the kHz QPO detections in the Sco-like Z sources are made \citep[see, e.g.,][]{vawiho1997,zhstsw1998,hovajo2002}.  The behavior of \sj\ in selections I--K is similar to that of the bright atoll sources GX 9+9, GX 9+1, and GX 3+1, for which no kHz QPOs have been reported either \citep[see, e.g.,][]{wivava1998}. Surprisingly, we did not detect kHz QPOs in the Cyg-like Z selections, despite the fact that all three persistent Cyg-like Z sources have shown kHz QPOs \citep{wihova1998,jowiva1998,jovaho2002}. As pointed out in \S \ref{sec:comparison}, however, selections A and B also show some differences from the Cyg-like Z sources in terms of CD/HID structure. Selection C is missing its horizontal branch, on which most kHz QPO detections were made in the Cyg-like Z sources. Finally, the fact that the Z-source-like and atoll-source-like kHz QPOs in \sj\ are well separated in luminosity and have distinct properties (such as Q-value and rms amplitudes) suggests that they might be the result of different excitation mechanisms, not only in \sj\, but also in other Z and atoll
sources.

\subsection{Long-term modulations}
 
Quasi-periodic long-term variability has been observed in numerous LMXBs \citep{chclco2008,ducore2010}. The long-term modulations seen in the low-energy light curve of \sj\ are closely related to the gradual changes in the CD/HID tracks of the source, and we therefore suggest that they are the result of changes in \mdot. They are only seen at intermediate luminosities (not at the peak of the outburst or during the decay). To study the time dependence of the long-term modulations we created a dynamical power spectrum of the long-term low-energy light curve \citep[see, e.g.,][]{clchco2003}, by combining Lomb-Scargle periodograms \citep{sc1982} of 75-day intervals, with the start dates of the intervals shifting by five days. The resulting dynamic power spectrum reveals that the period of the long-term modulations varies between $\sim$20--50 days. Given the strong variations in their period, the long-term variations are probably not the result of orbital modulations. We also note that the sharp dips that are present near the maxima of the first few cycles are not related to the absorption events that are often seen in high-inclination LMXBs; here they are the result of spectral changes associated with the source entering the horizontal branch.

The lack of a known orbital period for \sj\ complicates the interpretation of the long-term modulations, although, given the brightness and duration of the outburst, the orbital period of \sj\ is likely to be on the order of days or longer (H07). 
Accretion disk precession as the result of radiation-induced warping has been suggested as a source for long-term modulation in some LMXBs \citep{pr1996,wipr1999,ogdu2001}, although such modulations are usually discussed in terms of changes in viewing angle or obscuration (which we rule out as a source of subclass transitions in \sj - see \S\ref{sec:nature}). Although not explicitly discussed in these works, a precessing warped disc could perhaps also result in actual modulations in \mdot, as the result of varying torques from the secondary.

Two NS LMXBs have shown modulations that may be relevant to the discussion of the long-term variations in \sj. \citet{shbich2005} report on modulations with a period of $\sim$40 days in the NS LMXB 4U 1636--53. These modulations are accompanied by transitions between the island state and lower banana branch \citep[see also][]{behomo2007} and are therefore likely the result of modulations in \mdot. Similar to \sj, the modulations in 4U 1636--53 were not present at every luminosity level; they only appeared after the long-term average luminosity of the source had shown a decrease. \citet{shbich2005} argue that such a decrease could lower the irradiation of the outer accretion disc, resulting in a drop in its temperature and possibly bringing it into a regime of unstable accretion that results in limit cycle behavior. Based on the length and luminosity of its outburst, \sj\ is expected to have an orbital period on the order of days or more (H07), which is longer than 4U 1636--53 (3.8 hrs). Presumably, it therefore also has a larger accretion disc, which means that the onset of such an instability could already occur at a higher luminosity, in line with what we observe for \sj.

The long-term modulations in Cyg X-2 \citep{wikusm1996,pakima2000} are accompanied by strong changes  between various types of Z tracks \citep{kuvava1996}, similar to what we observe in \sj. Although the  modulations in Cyg X-2 appear to be less regular than in \sj, \citet{bosm2004} find that the lengths of the modulations cycles in Cyg X-2  are consistent with being
integer multiples (1--14) of the 9.843 day orbital period of the system \citep{cocrhu1979}.

\section{Summary \& Conclusions}

The outburst of \sj\ has revealed in great detail the evolution of a low-magnetic field NS-LMXB from near- or super-Eddington rates all the way down to quiescence. Most interestingly, \sj\ was observed to transform from a Cyg-like Z source to a Sco-like Z source at high luminosities, and from Sco-like Z to atoll source at low luminosties. A brief comparison with other NS-LMXBs suggests that the behavior observed in \sj\ is representative of the entire class of low-magnetic field NS-LMXBs.  We conclude that these transformations between NS-LMXB subclasses are the result of changes in a mass accretion rate component \mdot\ (which for most of the outburst is measured by
the 2--2.9 keV count rate) and argue that differences in neutron star spin/mass/magnetic field or viewing angle are of  minor importance to the Z/atoll phenomenology. The evolution observed within the Z source phase of the outburst shows a surprising evolution of the Z source tracks, with the horizontal, normal, and flaring branches disappearing one-by-one as \mdot\ appears to decrease. This is at odds with scenarios in which the same \mdot\ component increases or decreases monotonically from the horizontal to the flaring branch. We suggest that the \mdot\ responsible for the subclasses does not vary significantly along the Z tracks, while it does along atoll tracks. Motion along Z tracks may instead be the result of instabilities, whose presence {\it does} depend on \mdot.

\acknowledgments
We thank Andrea Sanna for pointing out a missed kHz QPO detection during the atoll phase of the outburst. TB acknowledges support from ASI grant ASI I/088/06/0. This research has made use of data obtained from the High Energy Astrophysics Science Archive Research Center (HEASARC), provided by NASA's Goddard Space Flight Center. 


\end{document}